\documentclass[english,aps,prl,reprint]{revtex4-1}
\pdfoutput=1
\usepackage[T1]{fontenc}
\usepackage[latin9]{inputenc}
\usepackage{amsbsy}
\usepackage{amssymb}
\usepackage{graphicx}
\usepackage{esint}

\makeatletter
 
 \@ifundefined{textcolor}{}
 {%
   \definecolor{BLACK}{gray}{0}
   \definecolor{WHITE}{gray}{1}
   \definecolor{RED}{rgb}{1,0,0}
   \definecolor{GREEN}{rgb}{0,1,0}
   \definecolor{BLUE}{rgb}{0,0,1}
   \definecolor{CYAN}{cmyk}{1,0,0,0}
   \definecolor{MAGENTA}{cmyk}{0,1,0,0}
   \definecolor{YELLOW}{cmyk}{0,0,1,0}
 }

\makeatother

\usepackage{babel}
\begin{document}

\title{Efficient Langevin Simulation of Coupled Classical Fields and Fermions}

\author{Kipton Barros}

\email{kbarros@lanl.gov}

\selectlanguage{english}%

\author{Yasuyuki Kato}

\affiliation{Theoretical Division and CNLS, Los Alamos National Laboratory, Los
Alamos, NM 87544}
\begin{abstract}
We introduce an efficient Langevin method to study bilinear fermionic
Hamiltonians interacting with classical fields.
Our approach is orders of magnitude faster than previous methods when applied
to very large systems with high accuracy requirements. To demonstrate the
method, we study complex non-coplanar chiral spin textures on the
triangular Kondo lattice model. We also explore non-equilibrium mesoscale
physics such as chiral domain coarsening and $\mathbb{Z}_{2}$ vortex
annihilation.
\end{abstract}
\maketitle
\global\long\def\mathd{\mathrm{{d}}}
\global\long\def\tr{\mbox{{Tr}}}

Lattice models of fermions interacting with classical fields encompass
a wide range of physics. Popular examples in condensed matter include Kondo
lattice (KL) models of  itinerant electrons interacting with localized magnetic moments~\cite{Doniach77}, Falicov-Kimball models of metal-insulator transitions in rare-earth materials~\cite{Falicov69} and Bogoliubov-De Gennes equations for superconductivity~\cite{De-Gennes66}. The Hubbard-Stratonovich transformation is another path to obtaining bilinear fermionic systems coupled to an auxiliary classical field~\cite{Blankenbecler81,Hirsch86}. This broad class of models poses a notoriously difficult numerical challenge: Monte Carlo (MC) sampling of the classical field requires \emph{repeated} diagonalization of the single-particle fermion matrix.

Several MC methods have been developed to more efficiently sample the classical field~\cite{Alonso01,Furukawa04,Alvarez07,Weisse09}. Spurred by colossal magnetoresistance
(CMR)~\cite{Jonker50,Ramirez97,Dagotto01}, these methods have largely been applied to the ferromagnetic transition in KL models at large coupling. This transition is relatively easy to study using moderate temperatures and small system sizes.

Recent interest has shifted to exotic spin-textures, which would occur in KL models at small to moderate couplings.  Skyrmion lattices have recently been observed with spatial modulations up to 0.1$\mu$m~\cite{Yu10,Solenov12}. Chiral textures lead to an anomalous Hall effect associated with huge ($\sim10^{5}T$) effective magnetic fields, as predicted in the KL model with triangular lattice~\cite{Martin08,Kato10} and experimentally observed in Pr$_{2}$Ir$_{2}$O$_{7}$~\cite{Machida10} and UCu$_{5}$~\cite{Ueland12}. Compared to ferromagnetism, these spin-textures can be very challenging to study. High precision and large system sizes may be needed to capture the physics of low temperatures and effective long-range interactions. State of the art numerical methods are often impractical.

In this Letter we introduce a suitable Langevin sampling method that is very
efficient---the cost scales \emph{linearly} with system size---at
high accuracy. Our method is based on a non-trivial gradient transformation~\cite{Griewank89}
of the kernel polynomial method (KPM)~\cite{Silver94}. Below we
outline our method and then demonstrate it with a study of the triangular KL model. Our lattices are large enough
to uncover interesting non-equilibrium effects such as chiral domain
coarsening and $\mathbb{Z}_{2}$ vortex dynamics. In this way, we
bridge the gap between quantum atomic scale and mesoscale physics.

Our method applies to a general bilinear fermionic Hamiltonian coupled to continuous, classical degrees of freedom $\phi$,
\begin{equation}
\mathcal{H}=\sum_{ij} c_{i}^{\dagger}A_{ij}(\phi)c_{j},\label{eq:hamiltonian}
\end{equation}
with sparse matrix $A$. We work at fixed temperature $\beta^{-1}$
and chemical potential $\mu$. The partition function is a trace over
classical and fermionic degrees of freedom, $Z=\tr_{\phi}\tr_{c}\exp[-\beta(\mathcal{H}-\mu\sum_{i}c_{i}^{\dagger}c_{i})]$.
Evaluating the fermionic trace yields $Z=\tr_{\phi}\exp(-\beta F)$,
where 
\begin{equation}
F(\phi)=\int\rho(\epsilon)f(\epsilon)\mathd\epsilon\label{eq:freeenergy}
\end{equation}
is the effective (free) energy of configuration $\phi$, $\rho(\epsilon)=\sum_{\nu}\delta(\epsilon-\epsilon_{\nu}(\phi))$
is the density of states of $A(\phi)$, and $f(\epsilon)=-\beta^{-1}\log\{1+\exp[-\beta(\epsilon-\mu)]\}$.
The energy contains effective long-range many-body interactions of
the classical field.

A key difficulty in MC sampling the classical field is the calculation
of $\Delta F$ in response to changes in $\phi$. KPM estimates the
density of states $\rho(\epsilon)$ using a series of Chebyshev polynomials
$T_{m}(\epsilon)$ truncated at order $M$~\cite{Silver94,Weisse06}.
The cost of KPM is linear in system size $N$ for sparse matrix $A$.
We state directly the recursive KPM procedure to construct an unbiased
stochastic estimate of $F(\phi$),

\begin{eqnarray}
F & = & \sum_{m=0}^{M-1}C_{m}\mu_{m}\label{eq:f_rec}\\
\mu_{m} & = & r^{\dagger}\cdot\alpha_{m}\\
\alpha_{m} & = & \left\{ \begin{array}{ll}
r & m=0\\
Ar & m=1\\
2A\alpha_{m-1}-\alpha_{m-2} & m>1
\end{array}\right.\label{eq:alpha_rec}
\end{eqnarray}
Here, $r$ is a random column vector whose components satisfy $\langle r_{i}^{*}r_{j}\rangle=\delta_{ij}$.
We draw complex $r_{i}$ from a uniform distribution $|r_{i}|^{2}=1$. The coefficients
\[
C_{m}=\int_{-1}^{1} \left(\pi\sqrt{1-\epsilon^{2}}\right)^{-1}\left(2-\delta_{0,m}\right)g_{m}T_{m}\left(\epsilon\right)f(\epsilon)\mathd\epsilon
\]
are independent of $A$. The Jackson kernel
\[
g_{m}=\frac{\left(M-m+1\right)\cos\frac{\pi m}{M+1}+\sin\frac{\pi m}{M+1}\cot\frac{\pi}{M+1}}{M+1}
\]
is chosen to damp Gibbs oscillations yet retain high accuracy. KPM
requires that the eigenvalues of $A$ have magnitude less than 1,
which can usually be achieved without loss of generality as a rescaling
of energy.

There are two independent sources of error in KPM: (1) truncation
at finite order $M$ and (2) stochastic estimation by averaging over
finitely many random vectors $r$. Both are well controlled and will
be discussed below.

To sample fields $\{\phi\}$ from the Boltzmann distribution, $P[\phi]\propto\exp(-\beta F[\phi]$),
we apply the overdamped Langevin equation. In discretized form,
\begin{equation}
\phi_{i}(t+\Delta t)-\phi_{i}(t)=-\Delta t\frac{\partial F}{\partial\phi_{i}}+\sqrt{2\beta^{-1}\Delta t}\eta_{i}(t),\label{eq:langevin}
\end{equation}
where $\eta_{i}(t)$ are uncorrelated Gaussian random variables with
unit variance and $t$ is a fictitious time. The Langevin approach
simultaneously updates all components $\phi_{i}$. Efficient and accurate
estimation of the gradient $\partial F/\partial\phi_{i}$
is crucial. We exclude inertial terms from the Langevin equation because
they would amplify errors in the gradient estimate.

The technique of automatic differentiation with ``reverse accumulation''~\cite{Griewank89}
ensures that, by careful application of the chain rule, we can transform
the KPM procedure to estimate $F$, Eqs.~\ref{eq:f_rec}--\ref{eq:alpha_rec},
into one that estimates $\partial F/ \partial \phi_i$ at the \emph{same cost}. We perform
this transformation analytically and state only the final result,

\begin{equation}
\frac{\partial F}{\partial A_{ij}}=\beta_{0;i}\alpha_{0;j}+2\sum_{m=1}^{M-2}\beta_{m;i}\alpha_{m;j}\label{eq:dfdh2}
\end{equation}
The row vectors $\beta_{m}$ are given by \emph{reverse} recursion,
from $m=M-2$ down to $m=0$,
\begin{equation}
\beta_{m} = C_{m+1}r^{\dagger}+2\beta_{m+1}A-\beta_{m+2} \label{eq:betas-1}
\end{equation}
with $\beta_{m \geq M-1} = 0$. The desired gradient is $\partial F/\partial\phi_{i}=\sum_{kl}(\partial F/\partial A_{kl})(\partial A_{kl}/\partial\phi_{i})$.
The sequence of vectors $\alpha_{m}$ are the same as in the original
KPM, but are here required in reverse order. We recalculate them as
needed using $\alpha_{m}=2A\alpha_{m+1}-\alpha_{m+2}$. The recursion
begins with $\alpha_{M-1}$ and $\alpha_{M-2}$, which are available
at the end of the original KPM procedure.

Note that the procedure to estimate all components of $\partial F/\partial \phi_i$, Eqs.~\ref{eq:dfdh2}--\ref{eq:betas-1},
has computational cost equivalent to the original KPM procedure to
calculate $F$.

The gradient calculation also inherits the approximation errors of
KPM, controlled by two parameters: the truncation order $M$ of the
Chebyshev series, and the dynamical stochastic error $z \equiv \Delta t / Q$, where $Q$ is the number of of KPM random vectors used per time step. The KPM
estimated density of states $\rho(\epsilon)$ is resolved to order
$\Delta\epsilon/M$, where $\Delta\epsilon=\epsilon_{\mathrm{\mathrm{max}}}-\epsilon_{\mathrm{min}}$
is the span of extremal eigenvalues. The parameter $M$ should be
chosen large enough to resolve the physically relevant features of
the density of states. Finite stochastic error $z>0$ acts much like
an additional noise term in the Langevin dynamics, effectively rescaling
its magnitude by an amount $T\rightarrow T+\Delta T_{\mathrm{eff}}$.
For matrices of the form $A(J \phi)$, where $J \ll 1$ is a small coupling constant, the estimate of the Langevin force term $\Delta t \partial F/ \partial \phi_i$ includes a stochastic error that scales like $\Delta t J / \sqrt{Q} = \sqrt {J^2 z \Delta t}$. Comparison with Eq.~\ref{eq:langevin} suggests modeling this stochastic error as an additional Langevin noise term with a temperature that scales as  $\Delta T_{\mathrm{eff}} \sim J^{2}z$.
The parameter $z$ should be chosen small enough that $\Delta T_{\mathrm{eff}}\ll T$
for the smallest relevant temperature scale $T$.

Our Langevin sampling remains efficient at high accuracy: the cost
to integrate the Langevin equation one unit of time is $\mathcal{O}(NM/z)$.
To compare to Metropolis MC with local updates, we assume that one
unit of Langevin integration time roughly corresponds to a full MC
sweep in which all $N$ lattice sites are visited. Trial MC changes $\Delta\phi_{i}$
are accepted with a probability that depends on the change in energy,
$\Delta F$. Brute force exact diagonalization of matrix $A$ requires
$\mathcal{O}(N^{3})$ operations, and a full MC sweep costs $\mathcal{O}(N^{4})$.
This may be reduced to $\mathcal{O}(N^{3})$ by tracking the response
of the spectrum to low-rank changes in $A$~\cite{Golub96,Alvarez07}.
Further acceleration is possible with KPM approximation. A non-stochastic
Green's function method reduces the cost of a full MC sweep to $\mathcal{O}(MN^{2})$~\cite{Weisse09}.
In an alternative approach, if $A$ contains only local coupling in
$d$ dimensions, a full MC sweep may be performed at cost $\mathcal{O}(M^{d+1}N)$~\cite{Furukawa04}.
In simulations presented below with $d=2$, $N=100^{2}$, $M=1000$,
and $z=0.02$, our Langevin approach outperforms existing
methods by 2 orders of magnitude or more. Another linear cost algorithm
is based on hybrid Monte-Carlo~\cite{Alonso01}, but a direct comparison
is difficult because the cost of precision at low temperatures is
unclear.

\begin{figure}
\includegraphics[scale=0.5]{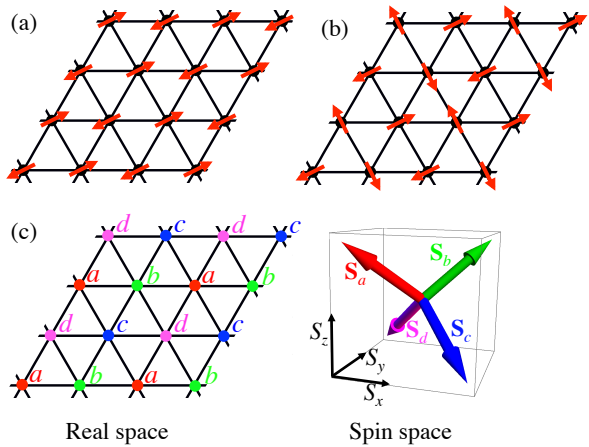}\caption{\label{fig:config}Three competing periodic $2\times2$ spin-textures
in the triangular Kondo lattice model at $3/4$ electron filling fraction.
The (a) $1q$, (b) $2q$, and (c) $3q$ (``all-out'') phases are
named according to their number of reciprocal lattice vectors. The
$3q$ phase maximizes chirality $\chi= \mathbf{S}_{i}\times\mathbf{S}_{j}\cdot\mathbf{S}_{k} =\pm4/3^{3/2}$
averaged over triangular plaquettes $[ i j k]$ and gives rise to a quantum
Hall effect at 1/4 and $3/4$ fillings. The $1q$ and $2q$ phases
break rotational symmetry of the triangular lattice.}
\end{figure}

We apply our method to the triangular KL model defined by the Hamiltonian,
\begin{equation}
\label{eq:kondo}
\mathcal{H}=-\sum_{ij\sigma}t_{ij}c_{i\sigma}^{\dagger}c_{j\sigma}-J\sum_{j\mu\nu}\mathbf{S}_{j}\cdot c_{j\mu}^{\dagger}\boldsymbol{\sigma}_{\mu\nu}c_{j\nu},
\end{equation}
where $c_{j\sigma}^{\dagger}$($c_{j\sigma}$) is the creation (annihilation)
operator of an electron with spin $\sigma$ on site $j$, $\mathbf{S}_{j}$
is a classical Heisenberg spin with $|\mathbf{S}_{j}|=1$, and $\mathbf{\boldsymbol{\sigma}_{\mu\nu}}=(\sigma_{\mu\nu}^{x},\sigma_{\mu\nu}^{y},\sigma_{\mu\nu}^{z})$
is a vector of Pauli matrices. The hopping coefficients are $t_{ij}=t$
when $i$ and $j$ are nearest neighbor sites on the triangular lattice, and $t_{ij}=0$
otherwise. In the following, we fix the energy scale by taking $t\rightarrow1$,
and the spatial scale by taking the lattice spacing to $1$.

Martin and Batista argue, by perfect nesting of the Fermi surface,
that the chiral $3q$ configuration (Fig.~\ref{fig:config}c) is
the ground state at 3/4 electron filling fraction with small coupling~\cite{Martin08}.
This state is of special interest, as it exhibits a spontaneous quantum
Hall effect. Variational calculation on the $2\times2$ plaquette
also predicts stability of the $3q$ state~\cite{Akagi10}. However,
unconstrained Monte-Carlo study of this phase at 3/4 filling has not yet been achieved
due to severe numerical difficulties. Very low temperatures, $T\lesssim0.001$
and small couplings $J\lesssim0.3$ are required to stabilize $3q$.
The numerical method must be very accurate to resolve
the small gap in the density of states (of width $\sim J^{2}$). Furthermore,
the $3q$ state is stabilized by a susceptibility that diverges like
$\log^{2}N$, so very large lattices sizes are required ($N\approx100^{2}$).
Due to these challenges, this system offers a rigorous test of
our Langevin method.

We choose $J=0.2$, $\mu=1.947$, and $N=100^{2}$. The three phases in
Fig.~\ref{fig:config} have very similar energy densities: $-4.15552,-4.15550,-4.15525$ for $3q$, $2q$, and $1q$, respectively. The ferromagnetic energy density, $-4.15278$, is not competitive.
We use KPM based Langevin sampling with $M=1000$ and
$z=0.02$. At $M=1000$, KPM estimates of energy differences are accurate
to order $10^{-5}$. With $z=0.02$ the effective Langevin temperature
is increased by $\Delta T_{\mathrm{eff}}\approx0.0002$.

\begin{figure}
\includegraphics{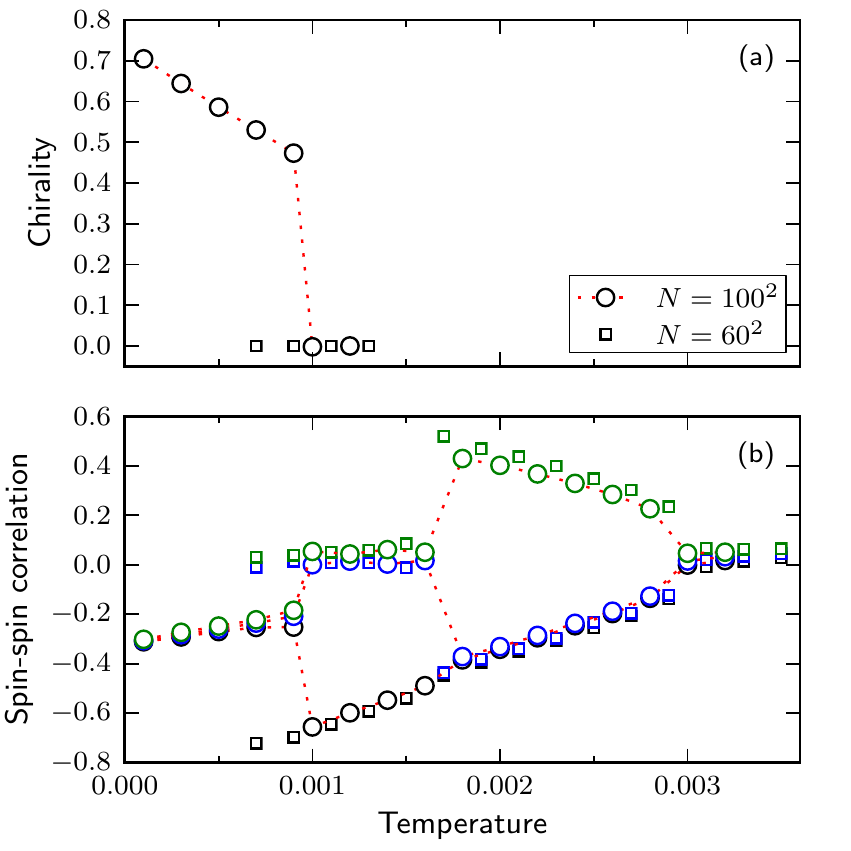}

\caption{\label{fig:phases}Phase diagram of the triangular Kondo lattice model
at Hund coupling $J=0.2$ and chemical potential $\mu=1.947$, corresponding
to filling fraction $\sim3/4$. (a) At low temperatures the preferred
$3q$ phase is identified by its non-zero mean chirality $\langle \chi \rangle$.
Very large system sizes are required to stabilize the $3q$ phase;
$N=100^{2}$ (circles) is sufficient, but $N=60^{2}$ (squares) is
not. (b) Spin-spin correlation functions $\langle\mathbf{S}_{i}\cdot\mathbf{S}_{j}\rangle$
for three nearest-neighbor orientations. Three first order phase transitions
are apparent: (1) $3q$ to $2q$ at $T=0.0010$, (2) $2q$ to $1q$
at $T=0.0017$, and (3) $1q$ to paramagnet at $T=0.0029$.}
\end{figure}
In Fig.~\ref{fig:phases}(a) we observe melting of the chiral $3q$
ground state. The mean chirality $\chi= \mathbf{S}_{i}\times\mathbf{S}_{j}\cdot\mathbf{S}_{k}$ of triangular plaquettes $[ijk]$ abruptly disappears at a first
order phase transition at $T\approx0.0010$. This transition, however,
is not to a paramagnetic phase. To distinguish the phases $3q$,
$2q$, and $1q$ we consider the (unordered) set of nearest-neighbor
spin-spin correlations $C=\{\langle\mathbf{S}_{\mathbf{x}}\cdot\mathbf{S}_{\mathbf{x}+\langle1,0\rangle}\rangle,\langle\mathbf{S}_{\mathbf{x}}\cdot\mathbf{S}_{\mathbf{x}+\langle1,\sqrt{3}\rangle/2}\rangle,\langle\mathbf{S}_{\mathbf{x}}\cdot\mathbf{S}_{\mathbf{x}+\langle-1,\sqrt{3}\rangle/2}\rangle\}$
averaged over lattice sites $\mathbf{x}$. Pure $3q$, $2q$, and
$1q$ phases would yield $C_{3q}=\{-1/3,-1/3,-1/3\}$, $C_{2q}=\{0,0,-1\}$,
and $C_{1q}=\{-1,-1,1\}$. The latter two states have broken bond
symmetry (specifically, the 3-fold rotational symmetry of the triangular
lattice).

Fig.~\ref{fig:phases}(b) plots the three elements of $C$ as a function
of temperature. At $T=0$ we find $C=C_{3q}$ as expected. We now
observe \emph{three} first order transitions at temperatures $T=0.0010$,
$0.0017$, and $T=0.0029$ to the $2q$, $1q$, and paramagnetic phases,
respectively. The $2q$ phase is identified by its correlation set
$C$, which has two zero elements and one negative element. In the
$1q$ phase, $C$ has one positive element and two negative (symmetric)
elements. To avoid equilibration issues in the above data, we used
initial conditions with explicitly broken chiral symmetry, $\langle \chi \rangle >0$, 

\begin{figure*}
\includegraphics[scale=0.5]{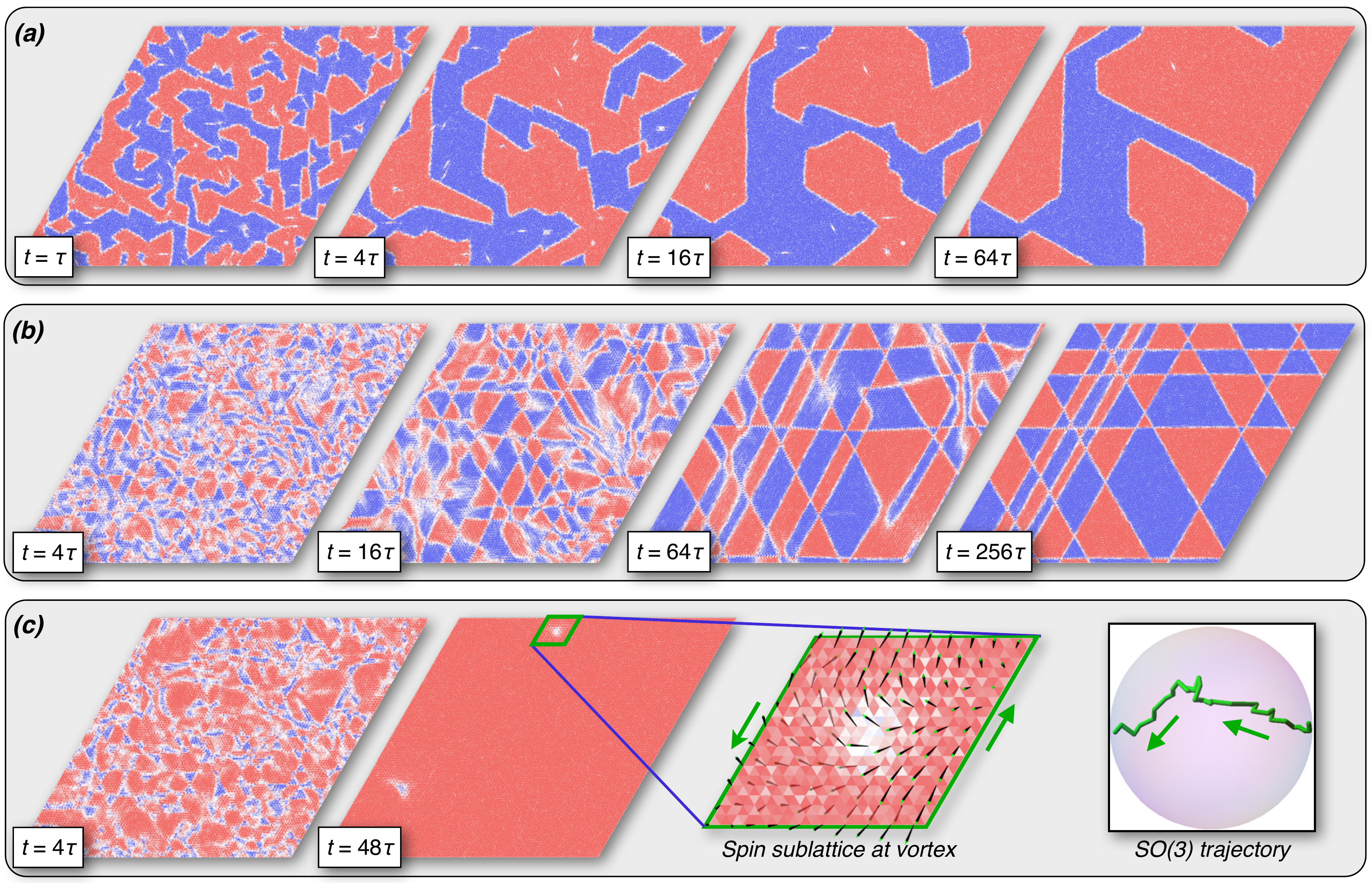}\caption{\label{fig:coarsening}Phase ordering in the $200\times200$ triangular
Kondo lattice model following a quench from infinite to zero temperature.
The color gradient, ranging from red to blue, is the local chirality.
Langevin times are measured in units of $\tau=1.28\times10^{4}$.
(a) $J=3$, $\mu=-3.2$ (\emph{$\sim1/4$} filling). Domain coarsening
with strong anisotropy is observed. $\mathbb{Z}_{2}$ vortices appearing
as white dots rapidly annihilate each other. (b) $J=0.2$, $\mu=1.947$
(\emph{$\sim3/4$} filling). The system is dynamically trapped in
a complex, robust metastable state. (c) $J=0.2$\emph{,} $\mu=1.947$,
$B_{z}=8\pi/\sqrt{3N}$ (\emph{$\sim3/4$} filling). An external field
breaks chiral symmetry and the system rapidly evolves to the $3q$
ground state. A $\mathbb{Z}_{2}$ vortex is identified by the winding
of a Burger's circuit (green) in $SO(3)$ space, a filled projective
sphere in the axis-angle representation.}
\end{figure*}

We now investigate the dynamical, non-equilibrium process by which
$3q$ chiral symmetry breaking occurs at low temperatures. We use
our Langevin dynamics to study the phase ordering kinetic of chiral
domains following a quench from infinite to zero temperature.

Langevin ``time'', properly speaking, is fictitious. However, in
the spirit of time-dependent Ginzburg-Landau (TDGL) models, we expect the energetic
relaxation of overdamped Langevin dynamics to qualitatively capture the large-scale aspects of phase
ordering~\cite{Chaikin00}. A point of comparison is Model A dynamics
in the Hohenberg and Halperin classification~\cite{Hohenberg77},
the prototypical TDGL model for phase-ordering of a non-conserved
scalar order parameter. Interestingly, we find
that the KL model has effective effective long-range many-body interactions that introduce dynamical features not present Model A.

First we consider the case of $\sim1/4$ filling with $J=3$ and $\mu=-3.2$.
Previous work found a robust $3q$ phase at system sizes up to $N=16^{2}$~\cite{Kato10}.
We use our Langevin dynamics to study the ordering dynamics at $N=200^{2}$
with accuracy parameters $M=500$, $z=0.005$. Figure~\ref{fig:coarsening}(a)
shows the evolution of chirality ranging from red (positive) to blue
(negative). The coarsening of chiral domains is analogous to Model
A, but we observe strong anisotropy of domain walls. The dynamics
slows at large times, consistent with a characteristic length scale
that grows as $\ell\sim t^{1/2}$~\cite{Bray94}.

Next we consider $\sim3/4$ filling, with $J=0.2$ and $\mu=1.947$.
We use accuracy parameters $M=500$ and $z=0.02$. The ordering dynamics
is displayed in Fig.~\ref{fig:coarsening}(b). A new dynamical feature
appears: the chiral domains evolve into a remarkable pattern which
is a very long lived metastable state. This is a reproducible phenomenon.
At higher temperatures this metastable pattern could be annealed to
the pure $3q$ phase, but in experimental practice it is easier to
explicitly break chiral symmetry with an applied external magnetic field $B_{z}$~\cite{Machida10}.
We introduce orbital coupling into our model,  Eq.~\ref{eq:kondo}, by applying  a
non-uniform phase $\theta_{ij}=B_{z}\hat{z}\cdot\mathbf{x}_{i}\times\mathbf{x}_{j}/2$ to the hopping coefficients $t_{ij} = t \exp(-i\theta_{ij})$, with $\mathbf{x}_{i}$ the position of lattice site $i$. The smallest magnetic field consistent with periodic boundaries,
$B_{z}=8\pi/\sqrt{3N}$, causes the system to rapidly reach the
uniform chiral $3q$ phase, shown in Fig.~\ref{fig:coarsening}(c).

Topological defects, visible as small white dots, are apparent at
both $1/4$ and $3/4$ filling. These defects are $\mathbb{Z}_{2}$
vortices associated with winding of the $SO(3)$ topological manifold,
and predicted to have fractional charge~\cite{Muniz12}.
A $\mathbb{Z}_{2}$ vortex is enlarged in Fig.~\ref{fig:coarsening}(c),
second and third panels. One of the four $3q$ spin sub-lattices are
shown. We can understand this defect by constructing a closed Burger's
circuit (draw in green) that encircles it. Every point on the green
circuit is identified with an element in $SO(3)$, plotted as a trajectory
in the fourth panel. The $SO(3)$ manifold, in the axis-angle representation,
is a filled-sphere with antipodal points identified. This Burger's
circuit has winding number 1 because it wraps $SO(3)$. The vortex
is $\mathbb{Z}_{2}$ because the only homotopically distinct winding
numbers are 0 and 1. Consequently, any pair of $\mathbb{Z}_{2}$ vortices
may annihilate. In the ordering dynamics of Fig.~\ref{fig:coarsening}(a),
we observe many vortices annihilating with each other and with domain
walls.

In conclusion, we have introduced a numerical method to study the
broad class of Hamiltonians that couple fermions to classical degrees
of freedom. Our method is highly accurate and efficient, enabling
the study of complex systems at unprecedented size. Large system sizes
may be necessary to resolve logarithmic divergences associated with
nesting of the fermi surface. In the triangular Kondo lattice model
at $3/4$ filling, we found that lattices of size $N=100^{2}$ with
six digits of precision are required. Large system sizes also allow
us to bridge the gap between quantum and mesoscopic physics. With
systems of size $N=200^{2}$ we are able to probe chiral domain dynamics,
metastable trapping, and $\mathbb{Z}_{2}$ vortex dynamics, effects
typically inaccessible in standard approaches.

We thank Ivar Martin and Cristian Batista for useful discussions.
This work was carried out under the auspices of the NNSA of the U.S.
DOE at LANL under Contract No. DE-AC52-06NA25396 and supported by
the LANL/LDRD Program. The calculations presented were performed using
the CCS-7 Darwin cluster.

\bibliographystyle{apsrev4-1}
\bibliography{/Users/kbarros/Dropbox/Pdfs/bibtex/journals,/Users/kbarros/Dropbox/Pdfs/bibtex/materials/0materials}

\end{document}